\documentclass[11pt]{article}

\input epsf
\usepackage{latexsym}
\usepackage{amssymb}
\usepackage{amsmath}
\usepackage[dvips]{graphicx}
\usepackage{array}

\def\d3{^{(3)}\nabla}

\begin{document}

\centerline{\Large \bf CMB and matter power spectra from cross correlations}
\vskip 0.2cm
\centerline{\Large\bf of primordial curvature and magnetic fields }

\vskip 2 cm

\centerline{Kerstin E. Kunze
\footnote{E-mail: kkunze@usal.es} }

\vskip 0.3cm

\centerline{{\sl Departamento de F\'\i sica Fundamental} and {\sl IUFFyM},}
\centerline{{\sl Universidad de Salamanca,}}
\centerline{{\sl Plaza de la Merced s/n, 37008 Salamanca, Spain }}

\vskip 1.5cm

\centerline{\bf Abstract}
\vskip 0.5cm
\noindent
A complete numerical calculation of the temperature anisotropies  and the polarization 
of the cosmic microwave background (CMB) is presented for 
a non zero cross correlation of a stochastic magnetic field with the primordial curvature
perturbation. Such a cross correlation results, for example, if the magnetic field is generated during inflation
by coupling electrodynamics to a scalar field which is identified with the curvaton. 
For a nearly scale invariant magnetic field of 1 nG it is found that at low multipoles the 
contribution due to the cross correlation dominates over that of the pure magnetic mode.
A similar behaviour on large scales is found for the linear matter power spectrum.

\vskip 1cm

\section{Introduction}
\setcounter{equation}{0}

The presence of magnetic fields on very different scales has been confirmed by observations \cite{obsgal}. On galactic scales 
there is an abundance of magnetic field detections with typical field strengths in the $\mu$G range at present time.
Over recent years there have been claims  of truly cosmological magnetic fields, that are not associated with any virialized 
structures such as galaxies or clusters of galaxies \cite{obscosmag}. However, this has been challenged in 
\cite{arlen}.

There are many models to explain the origin of large scale magnetic fields, for reviews see, e.g., \cite{reviews}. In particular, in the early universe 
conditions could have been such as to naturally generate magnetic fields strong enough to seed a galactic dynamo. 
The subsequent amplification since galaxy formation then results in $\mu$G magnetic fields today. Inflationary models
seem to be especially attractive to generate the initial seed magnetic field since the correlation lengths can be very large.
The problem usually is with the magnetic field strength which in spatially flat models is too small to account for the initial magnetic
seed field due to the global conformal invariance of standard electrodynamics in these backgrounds. 
However, in spatially curved models this is not the case and thus not a problem \cite{tsagas} (see however, \cite{adr}). 

Therefore, when considering the standard flat $\Lambda$CDM model
conformal invariance of electrodynamics has to be broken which  in the simplest case is realized by coupling to a scalar field \cite{ratra}. 
However, there are possible problems with back reaction effects onto  inflation \cite{dmr,problemsMag}.
Other possibilities include coupling electrodynamics to curvature  \cite{tw} of which some models also have problems due to 
ghost instabilities  \cite{hcp}. There are models in which magnetic fields of cosmologically relevant strengths can be generated during inflation 
\cite{tw, kk-mag1}.
However, since in this case inflation is of  power-law type there is yet another constraint coming from the fact 
the curvature perturbation for the relevant parameters cannot be created during inflation. It is possible to use the simplest 
curvaton scenario to generate the curvature perturbation after inflation \cite{lw}. This, however, results in a further restriction of the parameter space
but still allows to generate magnetic fields  strong enough to act as seed fields for the galactic dynamo
\cite{kk-magcurv}.

In \cite{cmk} a particular model has been studied in which magnetic fields are generated during inflation by coupling electrodynamics to 
a scalar field which is not the inflaton. 
In this case it has been shown that there is a non trivial correlation between the scalar field perturbation and the magnetic field \cite{cmk}.
Moreover, it was pointed out that identifying the (spectator) scalar field with the curvaton determines the cross correlation of the magnetic 
field with the curvature perturbation. This is  the case to be considered here.
The aim is to calculate the anisotropies of the temperature and polarization of the cosmic microwave background (CMB) due to the cross correlation 
between the magnetic field and the primordial curvature perturbation. Upto now, in general, the effect of magnetic fields present before  decoupling 
on the CMB have been calculated assuming either complete correlation \cite{grant}-\cite{kk1} or no correlation at all \cite{le,kk1} between the 
primordial curvature perturbation and the magnetic field. In  most works the magnetic field is assumed to be non helical. However, the  helical 
case has also been studied which is particularly interesting since it induces odd parity correlations of the CMB modes \cite{helical}.
In \cite{evolv} the effect of a non vanishing cross correlation between the curvature and the magnetic field due the evolution of the magnetic field 
has been studied. In that case the effect on the CMB temperature anisotropies and polarization has been found to be much below that of the 
compensated mode for a magnetic field which only redshifts with the expansion of the universe.

The magnetic field does not enter linearly in the perturbation equations but rather in form of its energy density and 
anisotropic stress which are quadratic in the magnetic field implying that non gaussianity is induced even at linear order. 
The resulting bispectra have been calculated  \cite{magNG}.
Thereby the cross correlations between the curvature mode and the magnetic field contributions are determined 
by the bispectra of the magnetic field and the curvature mode.
In \cite{mc} the bispectra induced by  coupling of the inflaton to electromagnetism have been calculated. This model has been 
also considered in \cite{ssy} where the corresponding CMB anisotropies have been calculated.

\section{The model}
\setcounter{equation}{0}

The background metric is assumed to be of the form
\begin{eqnarray}
ds^2=a^2(\eta)\left(-d\eta^2+\delta_{ij}dx^idx^j\right),
\end{eqnarray}
where $a(\eta)$ is the corresponding scale factor.

\begin{figure}[h!]
\centerline{\epsfxsize=3.1in\epsfbox{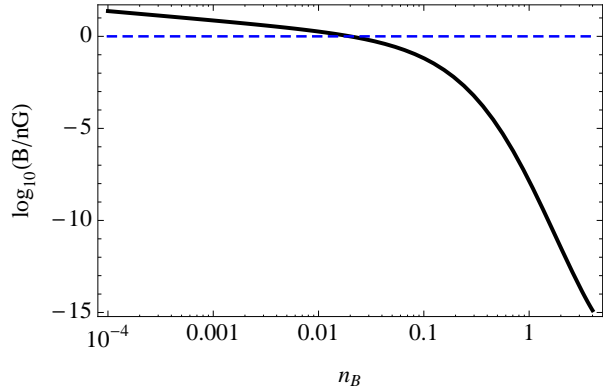}}
\caption{The root mean square magnetic field strength smoothed over the diffusion scale as a function of the spectral index $n_B$.}
\label{fig1}
\end{figure}

The model of magnetic field generation during de Sitter inflation used in \cite{cmk} which is based on \cite{dmr,by}  is described by the action
\begin{eqnarray}
S=\int d^4x \sqrt{-g}\left(-\frac{1}{4}W(\phi)F_{\mu\nu}F^{\mu\nu}-\frac{1}{2}\left(\partial\phi\right)^2-V(\phi)\right)
\end{eqnarray}
where the scalar field $\phi$ takes the role of a spectator field during inflation. Its potential is assumed to
be of the form $V(\phi)=-3nMH_I^2\phi$ and the coupling to the Maxwell tensor field, $F_{\mu\nu}$, is chosen to be
$W(\phi)=e^{2\phi/M}$. Moreover, $H_I=const.$ is the Hubble parameter and the scale factor is determined 
by $a(\eta)=-1/(\eta H_I)$  for $-\infty<\eta\leq\eta_I<0$ upto the end of inflation
at $\eta_I$.
The resulting magnetic field is non helical
and its power spectrum  at the end of inflation on superhorizon scales is given by \cite{cmk}
\begin{eqnarray}
P_B(k,\eta_I)\simeq \frac{\Gamma\left(\frac{5-n_B}{2}\right)^2}{\pi}\left(\frac{-k\eta_I}{2}\right)^{n_B-4}k\left(H_I\eta_I\right)^4,
\end{eqnarray}
where 
\begin{eqnarray}
n_B=\left\{
\begin{array}{lrrr}
4+2(n+1) &\hspace{2cm} n & < &-1,\\
4-2n &   n & \geq & 0.
\end{array}
\right.
\end{eqnarray}
In the numerical solutions the magnetic field values are taken at present time. Thus assuming that the magnetic field only redshifts with
expansion, so that $B_i\sim1/a^2$. Then the  two point function of the magnetic field,
\begin{eqnarray}
\langle B_i^*(\vec{k}, \eta)B_j(\vec{k}',\eta')\rangle = P_B(k,\eta)\delta_{\vec{k},\vec{k}'}\delta_{\eta,\eta'}\left(\delta_{ij}-\frac{k_ik_j}{k^2}\right)
\end{eqnarray}
 determines that the power spectrum scales as $a^{-4}$, so that 
at present time 
\newline
$a_0^4P_B(k,\eta_0)=(H_I\eta_I)^{-4}P_B(k,\eta_I)$.
Then together with the parameters chosen as in \cite{cmk}, $-\eta_I=10^{-24}$ Mpc for the end of inflation, $a_0=1$, and introducing a pivot scale $k_L=1$ Mpc$^{-1}$ for the magnetic field spectrum and a gaussian window function $W(k,k_m)$ to model damping of the magnetic field on  small scales thereby imposing an upper cut-off of the wave numbers, $k_m$, the spectral function at present time is given by
\begin{eqnarray}
P_B(k,\eta_0)=10^{97.2-24.3 n_B}\frac{\Gamma\left(\frac{5-n_B}{2}\right)^2}{\pi}\left(\frac{k_L}{1 {\rm Mpc}^{ -1}}\right)^{n_B-3}\left(\frac{k}{k_L}\right)^{n_B-3}
W(k,k_m)
{\rm Mpc}^{-1}
\label{PB}
\end{eqnarray}
where  $W(k,k_m)=\pi^{-3/2}e^{-\left(\frac{k}{k_m}\right)^2}$, so that $\int d^3k W(k,k_m)=k_m^3$. 
The maximal wave number is given by \cite{sb}
\begin{eqnarray}
k_{\rm m}\simeq 198.454\left(\frac{B}{\rm nG}\right)^{-1}{\rm Mpc}^{-1},
\label{km}
\end{eqnarray}
 for the values of the bestfit $\Lambda$CDM model of  WMAP9, $\Omega_b=0.02264h^{-2}$ and $h=0.7$ \cite{wmap9}.
Comparing the spectrum (\ref{PB}) with the form of the magnetic field spectra used in, e.g., in \cite{kk1} the spectral index there becomes $n_B\rightarrow n_B+3$, so that the scale invariant cases correspond here to $n_B=0$ and in \cite{kk1} to $n_{B*}=-3$.
Thus the magnetic field strength today smoothed over the diffusion scale is determined by 
\begin{eqnarray}
\langle \vec{B}^2(\vec{x},\eta_0)\rangle=10^{-19.8-24.3n_B}\Gamma^2\left(\frac{5-n_B}{2}\right)\Gamma\left(\frac{n_B}{2}\right)\left(\frac{k_m}{1 {\rm Mpc}^{-1}}\right)^{n_B}{\rm G}^2.
\end{eqnarray}
Defining the magnetic field strength $B\equiv\sqrt{\langle \vec{B}^2(\vec{x},\eta_0)\rangle}$ and using equation (\ref{km})  yields to
\begin{eqnarray}
\left(\frac{B}{\rm nG}\right)^{\frac{n_B}{2}+1}=10^{-0.9-11n_B}\Gamma\left(\frac{5-n_B}{2}\right)\Gamma^{\frac{1}{2}}\left(\frac{n_B}{2}\right),
\end{eqnarray}
which is shown in figure \ref{fig1}. Therefore for $n_B=0.02$ corresponding to $n_{B*}=-2.98$ the smoothed magnetic field strength is  of the order of 1 nG. 
The scale invariant case $n_B=0$ corresponds to $n=2$ or $n=-3$. As shown in \cite{cmk} only the  case  $n=2$ is allowed since in the other case 
strong back reactions would develop since the electromagnetic energy density dominates over the inflaton energy density.

\section{Cross correlation functions of primordial curvature and magnetic fields}
\setcounter{equation}{0}

In \cite{cmk} the 3-point cross correlation between the magnetic field energy density proportional to $B^2$and the fluctuations of the spectator field $\delta\phi$ 
has been calculated.
However, since the anisotropies of the CMB are determined by the magnetic energy density contrast $\Delta_B$ and the magnetic anisotropic stress
$\pi_B$ it is necessary to obtain the 3-point cross correlation functions allowing for arbitrary components of the magnetic field.
Starting with the expression found for the cross correlation of the gauge potential $A_i$ of the electromagnetic field and $\delta\phi$ \cite{cmk} the 
3-point cross correlation involving arbitrary components of the magnetic field is determined to be, in Fourier space,
\begin{eqnarray}
\langle \frac{\delta\phi(\vec{k}_1,\eta_I)}{M}B_i(\vec{k}_2,\eta_0)B_j(\vec{k}_3,\eta_0)\rangle
=-(2\pi)^3\delta^{(3)}(\vec{k}_1+\vec{k}_2+\vec{k}_3)\epsilon_{ilm}k_{2l}\epsilon_{jnq}k_{3n}U_{mq},
\end{eqnarray}
where \cite{cmk}
\begin{eqnarray}
U_{mq}=-\frac{\pi^2}{8}\left(\frac{H_I}{M}\right)^2\frac{1}{k_1^4}\left[\delta_{mq}I_1+\left(\hat{k}_2\cdot\hat{k}_3\delta_{mq}-\hat{k}_{2q}\hat{k}_{3m}\right)I_2\right],
\end{eqnarray}
where $I_1$ and $I_2$ are integrals depending, in particular, on the parameter $n$ which can be found in \cite{cmk}. The special case of $n=2$ 
will be presented in the notation used here below.
This corresponds to a scale invariant magnetic field spectrum which is the cosmologically relevant case \cite{cmk}. 

In order to relate the perturbations in the scalar field to the curvature perturbation the spectator field is identified with a curvaton field as suggested in \cite{cmk}.
In particular, we assume the simplest curvaton model \cite{lw} so that the  total curvature perturbation $\zeta$ is generated by the curvaton after inflation and that any other contributions to the curvature perturbation are negligible. Moreover, it is assumed that the curvaton dominates the energy density before its decay.
Then following \cite{lw}
\begin{eqnarray}
\frac{\delta\phi(\vec{k})}{M}=\frac{3}{2}\left(\frac{\phi_*}{M}\right)\zeta(\vec{k}),
\end{eqnarray}
where the asterisk denotes the epoch when the mode $k$ leaves the horizon during inflation assuming no nonlinear evolution on large scales \cite{svw}. 
Therefore calculating $\langle \zeta_{\vec{k}}^*\zeta_{\vec{k}'}\rangle$ at the pivot scale of WMAP9 $k_p=0.002$ Mpc$^{-1}$
yields to
\begin{eqnarray}
\left(\frac{H_I}{\phi_*}\right)=3\pi{\cal P}_{\zeta}^{\frac{1}{2}}(k_p)
\end{eqnarray}
where the dimensionless power spectrum of two random variables $X$ and $Y$,  ${\cal P}_{\langle XY\rangle}$, defined by
\begin{eqnarray}
\langle X_{\vec{k}}^*Y_{\vec{k}'}\rangle=\frac{2\pi^2}{k^3}{\cal P}_{\langle XY\rangle}(k)\delta_{\vec{k},\vec{k}'}.
\end{eqnarray}
has been used.

The magnetic energy density contrast and anisotropic stress are defined in terms of the photon energy density $\rho_{\gamma}$ and pressure $p_{\gamma}=\frac{1}{3}\rho_{\gamma}$. This is due to the weakness of  the magnetic field which does not contribute to the background quantities (for a different approach see \cite{tsagas1}).
The energy density contrast $\Delta_B$ in $k$-space is defined by (for details, e.g., cf. \cite{kk1})
\begin{eqnarray}
\Delta_B(\vec{k})=\frac{1}{2\rho_{\gamma 0}}\sum_{\vec{q}}B_i(\vec{q})B^i (\vec{k}-\vec{q}).
\end{eqnarray}
So that one of the contributions to the cross correlation between the primordial curvature mode and the magnetic mode is determined by
the two-point correlation function $\langle \zeta(\vec{k}') \Delta_B(\vec{k})\rangle$. 
Taking the continuum limit $\sum_{\vec{k}}\rightarrow\int\frac{d^3 k}{(2\pi)^3}$ and 
using the expressions for the integrals $I_1$ and $I_2$ for $n=2$ on superhorizon scales as given in \cite{cmk} yields to
\begin{eqnarray}
{\cal P}_{\langle \zeta\Delta_B\rangle}(k)&=&\frac{9}{16\pi^{\frac{9}{2}}}\frac{{\cal P}_{\zeta}^{\frac{1}{2}}(k_p)}{\rho_{\gamma 0}\eta_I^4}
\left(\frac{H_I}{M}\right)e^{-\frac{1}{2}\left(\frac{k}{k_m}\right)^2}
\nonumber\\
&\times&
\int_0^{\infty}dz z^3e^{-\left(\frac{k}{k_m}\right)^2z^2}
\int_{-1}^1dx e^{\left(\frac{k}{k_m}\right)^2zx}
\vartheta\left[2\frac{x-z}{\vartheta}J_1+\left[1+\left(\frac{x-z}{\vartheta}\right)^2\right]J_2\right],
\end{eqnarray}
where 
\begin{eqnarray}
J_1&\equiv&\frac{(1+\vartheta)(1+\vartheta+\vartheta^2)+2(1+\vartheta+\vartheta^2)z+2(1+\vartheta)z^2+z^3}
{z^3\vartheta^3(1+z+\vartheta)^2},\nonumber\\
J_2&\equiv&\left[-3(1+z+\vartheta)^2(\gamma+\ln(1+z+\vartheta))+(1+\vartheta)^2(3+3\vartheta+\vartheta^3)
\right.\nonumber\\
&&\left.
+(1+\vartheta)(9+6\vartheta+2\vartheta^3)z
+(9+6\vartheta+2\vartheta^2+2\vartheta^3)z^2
\right.\nonumber\\
&&\left.
+2(2+\vartheta+\vartheta^2)z^3+2(1+\vartheta)z^4+z^5\right]z^{-4}\vartheta^{-4}(1+z+\vartheta)^{-2}.
\end{eqnarray}
Moreover,
\begin{eqnarray}
x\equiv\frac{\vec{k}\cdot\vec{q}}{kq},\hspace{3cm}
z\equiv\frac{q}{k},\hspace{3cm}
\vartheta\equiv\vartheta(x,z)\equiv(1-2zx+z^2)^{\frac{1}{2}}
\end{eqnarray}

\begin{figure}[h!]
\centerline{\epsfxsize=2.2in\epsfbox{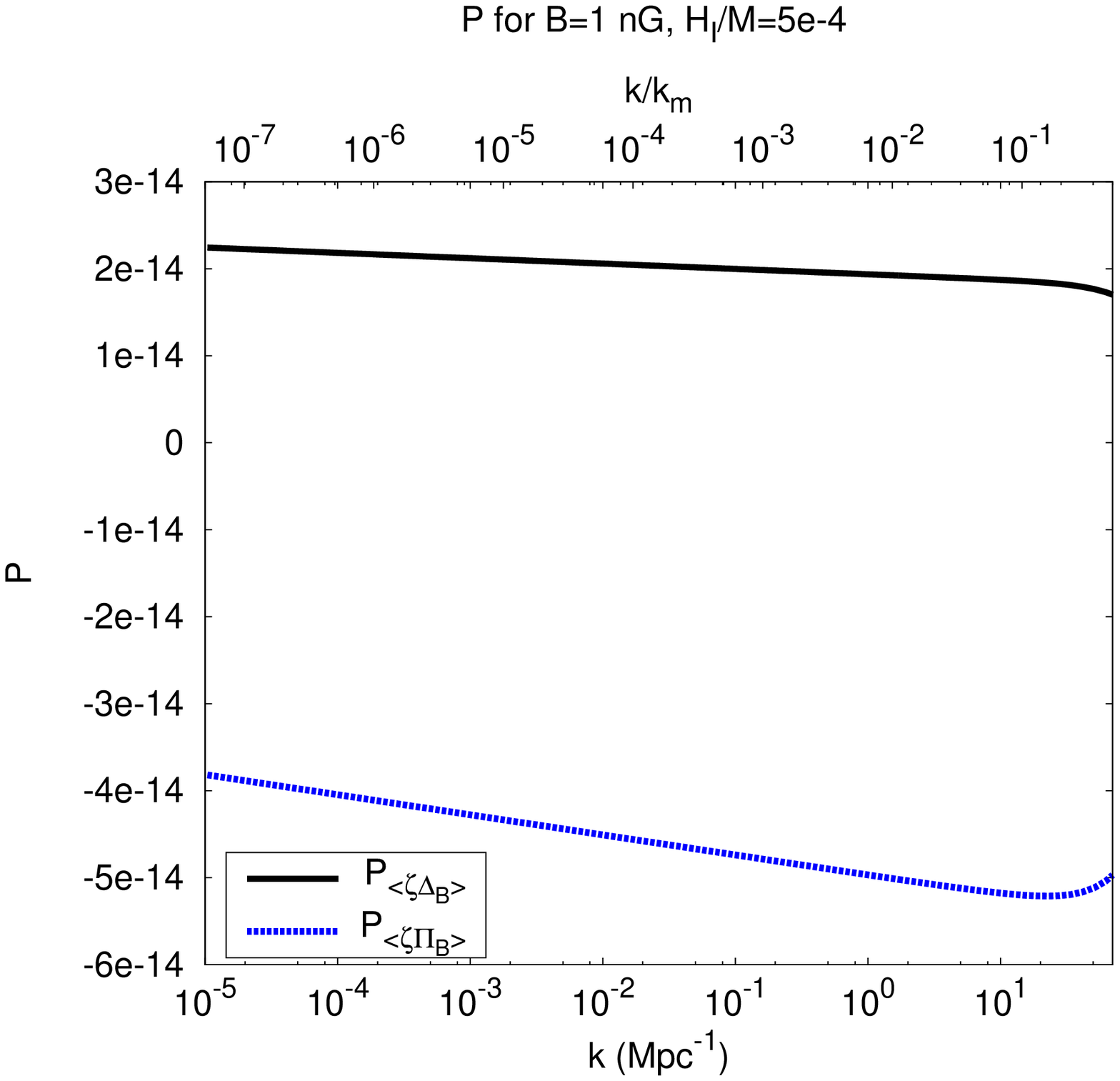}
\epsfxsize=2.2in\epsfbox{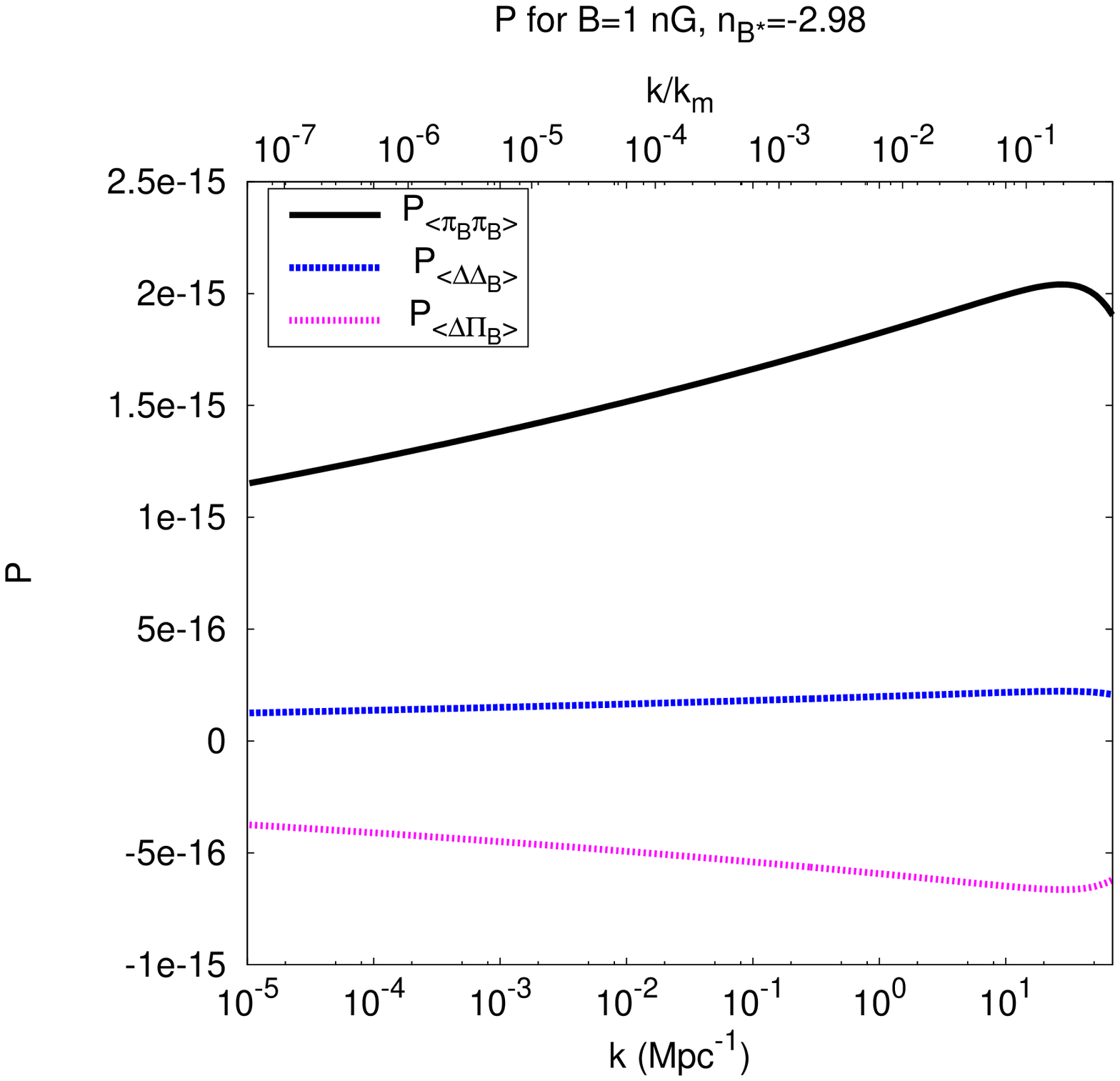}
\epsfxsize=2.2in\epsfbox{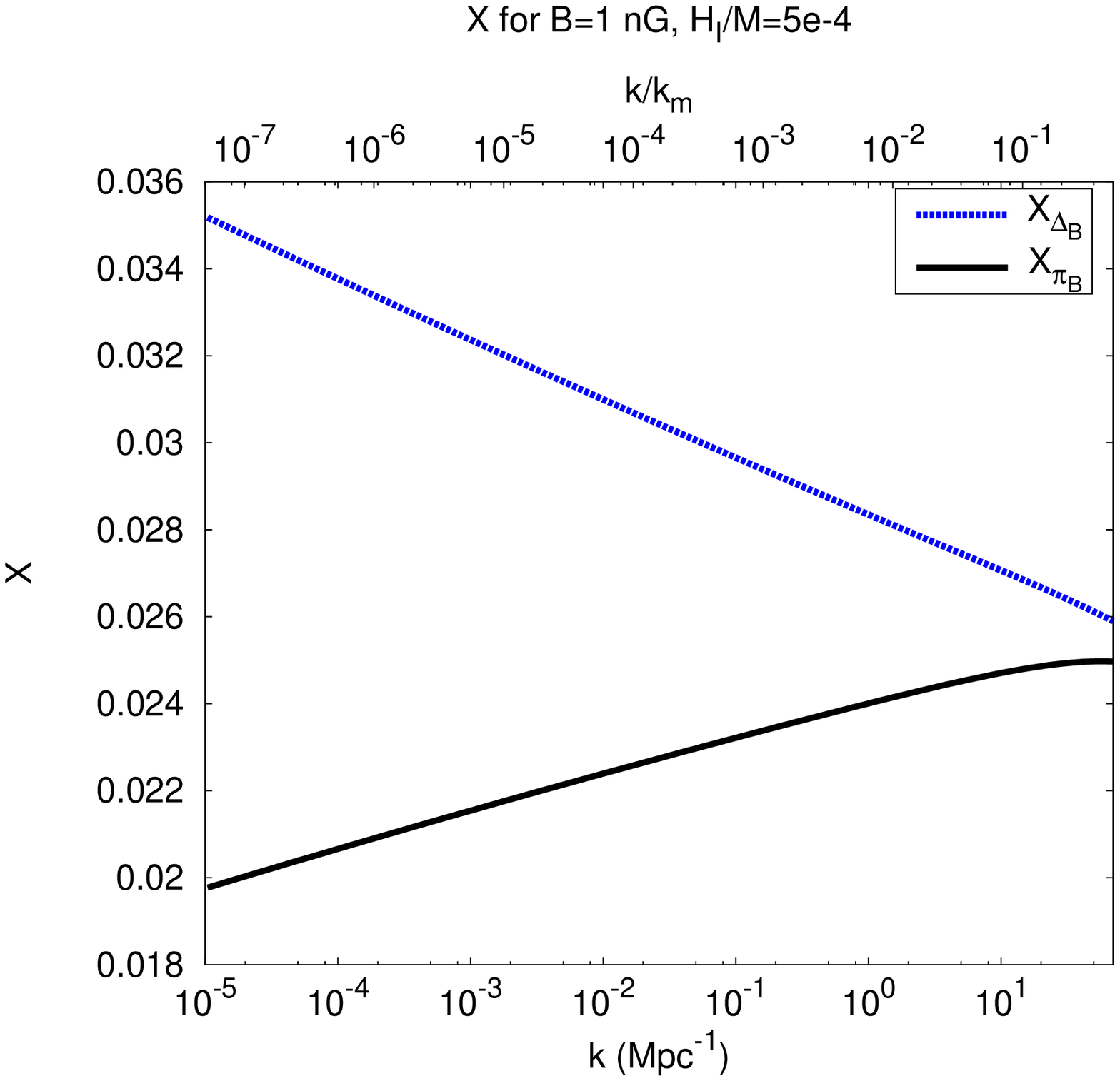}}
\caption{{\it Left:} The dimensionless spectra determining the cross correlation function of the primordial curvature perturbation and the magnetic energy density contrast as well as the magnetic anisotropic stress.
{\it Middle: }  The dimensionless spectra determining the auto- and cross correlations of the pure magnetic mode for $B=1$ nG for a nearly scale invariant magnetic field, which corresponds to $n_B=0.02$ or  $n_{B*}=-2.98$ in the conventions used, e.g.,  in \cite{kk1}.
{\it Right:} The ratios determining the generalized Schwarz inequality for $\langle \zeta\Delta_B\rangle$ and $\langle\zeta\pi_B\rangle$.
The numerical solutions are calculated for the parameters $\frac{H_I}{M}=5\times 10^{-4}$, $\eta_I=-10^{-24}$ Mpc  \cite{cmk}.}
\label{fig2}
\end{figure}

The two-point correlation function $\langle \zeta(\vec{k})\pi_B(\vec{k}')\rangle$ is calculated using the expression for the anisotropic stress
of the scalar mode (e.g., \cite{kk1})
\begin{eqnarray}
\pi_B(\vec{k})=\frac{3}{2\rho_{\gamma 0}}\left[\sum_{\vec{q}}\frac{3}{k^2}B_i(\vec{q})(k^i-q^i)B_j(\vec{k}-\vec{q})q^j-\sum_{\vec{q}}
B_m(\vec{q})B^m(\vec{k}-\vec{q})\right].
\end{eqnarray}
For the cosmologically interesting case $n=2$ this yields to
\begin{eqnarray}
{\cal P}_{\langle\zeta\pi_B\rangle}(k)&=&-\frac{27}{16\pi^{\frac{9}{2}}}\frac{{\cal P}_{\zeta}^{\frac{1}{2}}}{\rho_{\gamma 0}\eta_I^4}\left(\frac{H_I}{M}\right)
e^{-\frac{1}{2}\left(\frac{k}{k_m}\right)^2}
\nonumber\\
&\times&
\int_0^{\infty}dz z^3 e^{-\left(\frac{k}{k_{m}}\right)^2z^2}\int_{-1}^1 dx e^{\left(\frac{k}{k_m}\right)^2zx}
\left[\left(2x+(1-3x^2)z\right)J_1
\right.
\nonumber\\
&&\left.
+\left(\vartheta
+\frac{x-z}{\vartheta}\left[x+(2-3x^2)z\right]\right) J_2\right].
\end{eqnarray}
The dimensionless power spectra determining the cross correlation between the curvature mode and the magnetic modes  are shown in 
figure \ref{fig2} ({\it left}) for $\frac{H_I}{M}=5\times 10^{-4}$ corresponding to the upper bound on this parameter inferred in \cite{cmk}.
As can be appreciated from figure \ref{fig2}  whereas the curvature perturbation is correlated with  the magnetic energy density contrast, it is anticorrelated with the magnetic anisotropic stress. 
For comparison, in figure \ref{fig2} ({\it middle}) the auto- and cross correlations of the magnetic modes are shown for magnetic field strength
$B=1$ nG and spectral index $n_{B*}=-2.98$. The expressions for the correlation functions of the magnetic modes, ${\cal P}_{\langle\Delta_B\Delta_B\rangle}(k)$,
${\cal P}_{\langle\pi_B\pi_B\rangle}(k)$ and ${\cal P}_{\langle\Delta_B\pi_B\rangle}(k)$ can be found, e.g., in \cite{kk1}.
Comparing the {\it left} and {\it middle} figures of figure \ref{fig2} shows that the amplitudes of the 
spectral functions of the cross correlations between curvature and the magnetic modes are larger than the auto- and cross correlation
functions of the magnetic modes. In figure \ref{fig2} ({\it right}) the ratio 
\begin{eqnarray}
X_{\Xi}=\frac{|{\cal P}_{\langle \zeta\Xi\rangle}|}{\sqrt{|{\cal P}_{\zeta\zeta}||{\cal P}_{\langle \Xi\Xi\rangle}}|},
\hspace{4cm}
\Xi=\Delta_B, \pi_B 
\end{eqnarray}
is reported which determines the generalized Schwarz inequality and as can be appreciated is well below the unity bound.

\section{Results}
\setcounter{equation}{0}

The CMB angular power spectra are determined by the brightness function which in the line-of-sight approach \cite{sz} 
is written for each component, for the scalar mode, \cite{hw}
\begin{eqnarray}
\frac{\Theta_{\ell}^{X}(\eta_0,k)}{2\ell +1}=\int_0^{\eta_0}d\eta S_{\Theta}^{X}(k,\eta)j_{\ell}\left[k(\eta_0-\eta)\right]
\end{eqnarray}
where $S_{\Theta}^X$ is the source function and $X$ denotes $\zeta$, $\Delta_B$ and $\pi_B$. This determines the 
temperature auto correlation function 
\begin{eqnarray}
C_{\ell}^{TT, \langle XY \rangle}=\frac{1}{2\pi^2}\int_0^{\infty}\frac{dk}{k}\int_0^{\eta_0}d\eta{\cal P}_{\langle XY\rangle}(k)
S_{\Theta}^X(k,\eta)S_{\Theta}^Y(k,\eta)\left(j_{\ell}\left[k(\eta_0-\eta)\right]\right)^2
\end{eqnarray}
and similarly for the auto correlation function of the E-mode and the temperature polarization cross correlation.
The resulting angular power spectra are calculated using a modified version of CMBEASY \cite{cmbeasy}.
It is based on the modified version of \cite{kk1} where the initial conditions and evolution equations
including the magnetic field contributions can be found.
In all solutions the best fit values of the 6-parameter $\Lambda$CDM model of WMAP9 \cite{wmap9}
have been used, in particular, $\Omega_b=0.0463$, $\Omega_{\Lambda}=0.721$, ${\cal P}_{\zeta}(k_p)=2.41\times 10^{-9}$,
$n_s=0.972$ and the reionization optical depth $\tau=0.089$. Moreover, the nearly scale invariant magnetic field has field strength 
$B=1$ nG and spectral index $n_B*=-2.98$. The factor $\frac{H_I}{M}$ determining the cross correlation functions is 
$\frac{H_I}{M}=5\times 10^{-4}$.

\begin{figure}[h!]
\centerline{\epsfxsize=3in\epsfbox{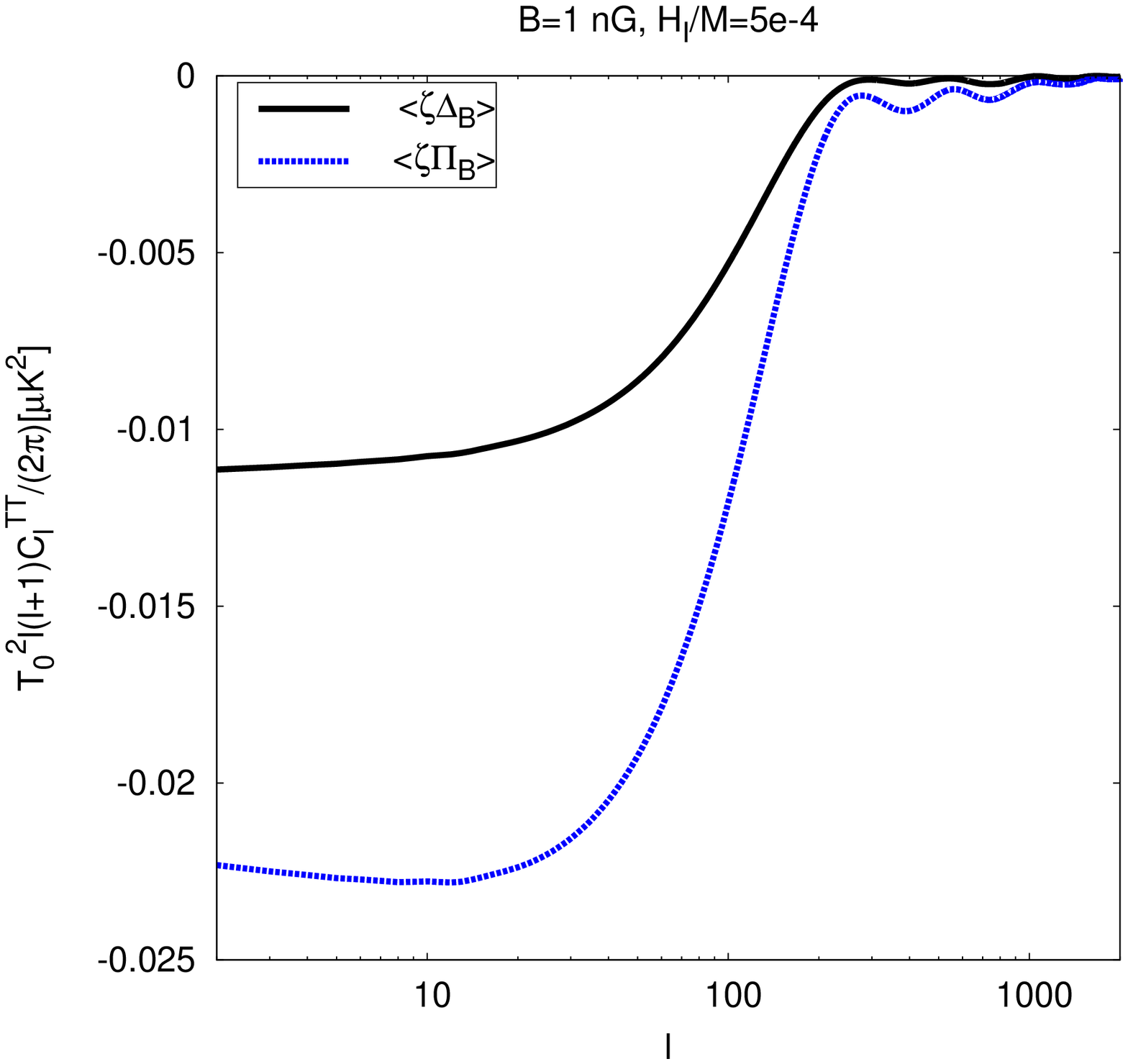}
\epsfxsize=3in\epsfbox{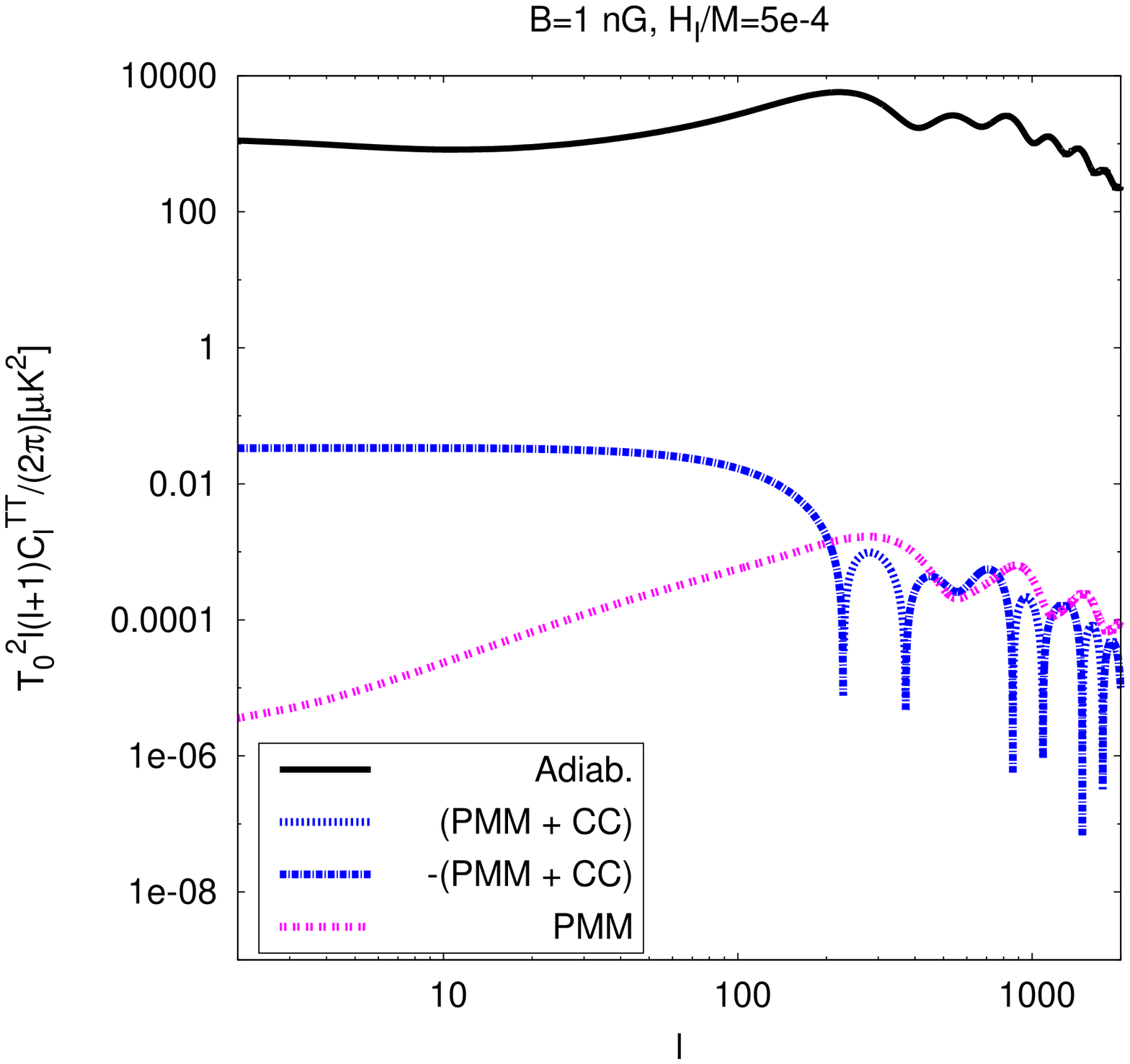}
}
\caption{Angular power spectra determining the temperature autocorrelation of the CMB
due to the cross correlation between the curvature mode $\zeta$ and the $\Delta_B$ mode 
 and due to the cross correlation between the curvature mode and the $\pi_B$ mode ({\it left}).
The contributions to the total angular power spectrum include the adiabatic mode ({\it Adiab.}) , the pure magnetic mode ({\it PMM}) and the 
total cross-correlated magnetic-curvature mode ({\it CC}) ({\it right}). These have been calculated for WMAP9 $\Lambda$CDM
best fit parameters.
}
\label{fig3}
\end{figure}

In figure \ref{fig3} the angular power spectra determining the temperature auto correlation of the CMB
are shown. 
The contribution due to the cross correlation between the curvature mode and the magnetic modes dominates  on large scales 
over the one due to the pure magnetic mode, that is the compensated mode for which the initial conditions are such that 
the contributions from the neutrino anisotropic stress and the magnetic anisotropic stress cancel each other. 
There is also the so called passive mode \cite{le} due to the presence of a magnetic field before neutrino decoupling which causes 
an additional contribution to the curvature perturbation amplitude proportional to the magnetic anisotropic stress. 
However, since this is an adiabatic like mode it is not further studied.
\begin{figure}[h!]
\centerline{\epsfxsize=3in\epsfbox{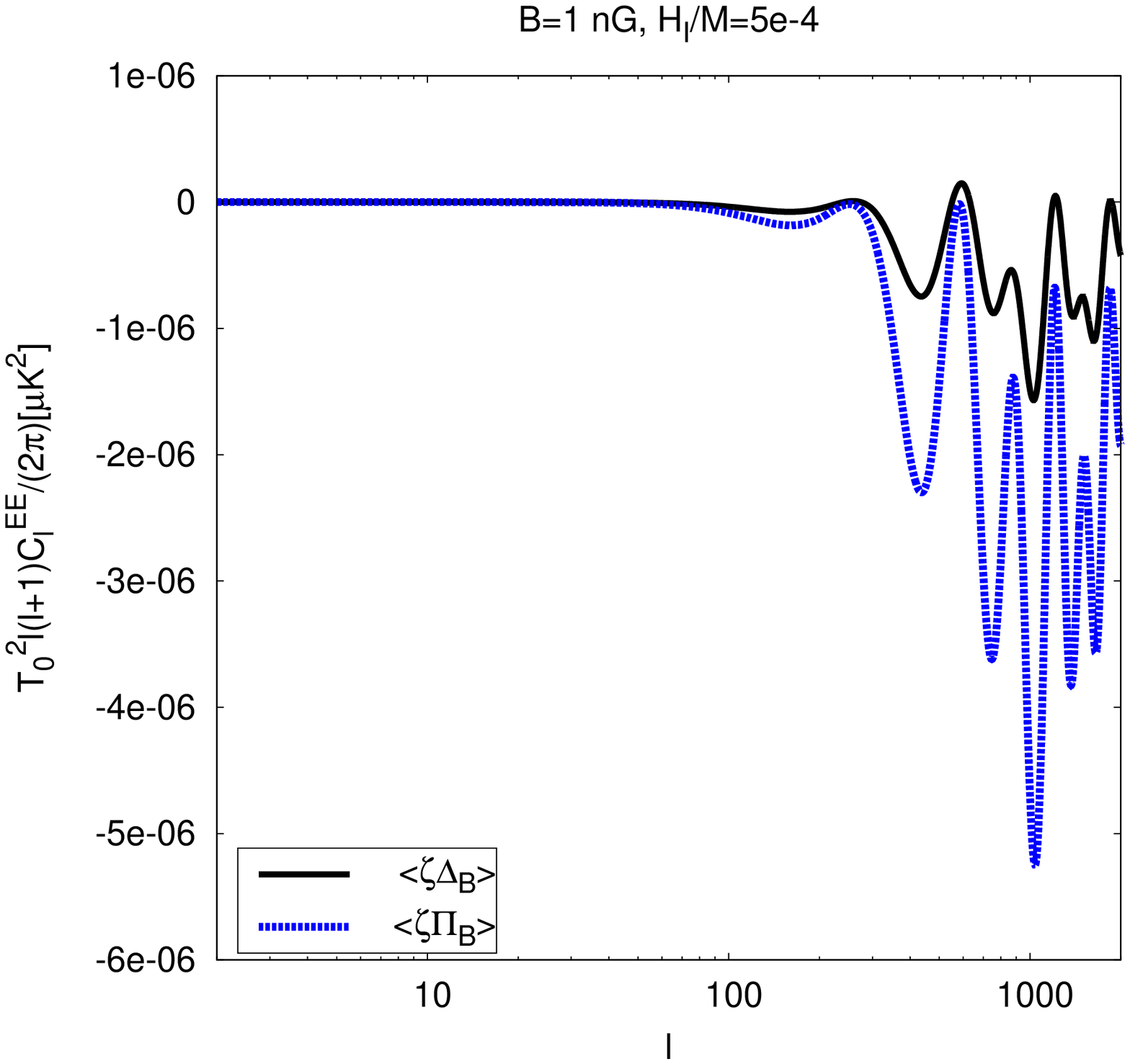}
\epsfxsize=3in\epsfbox{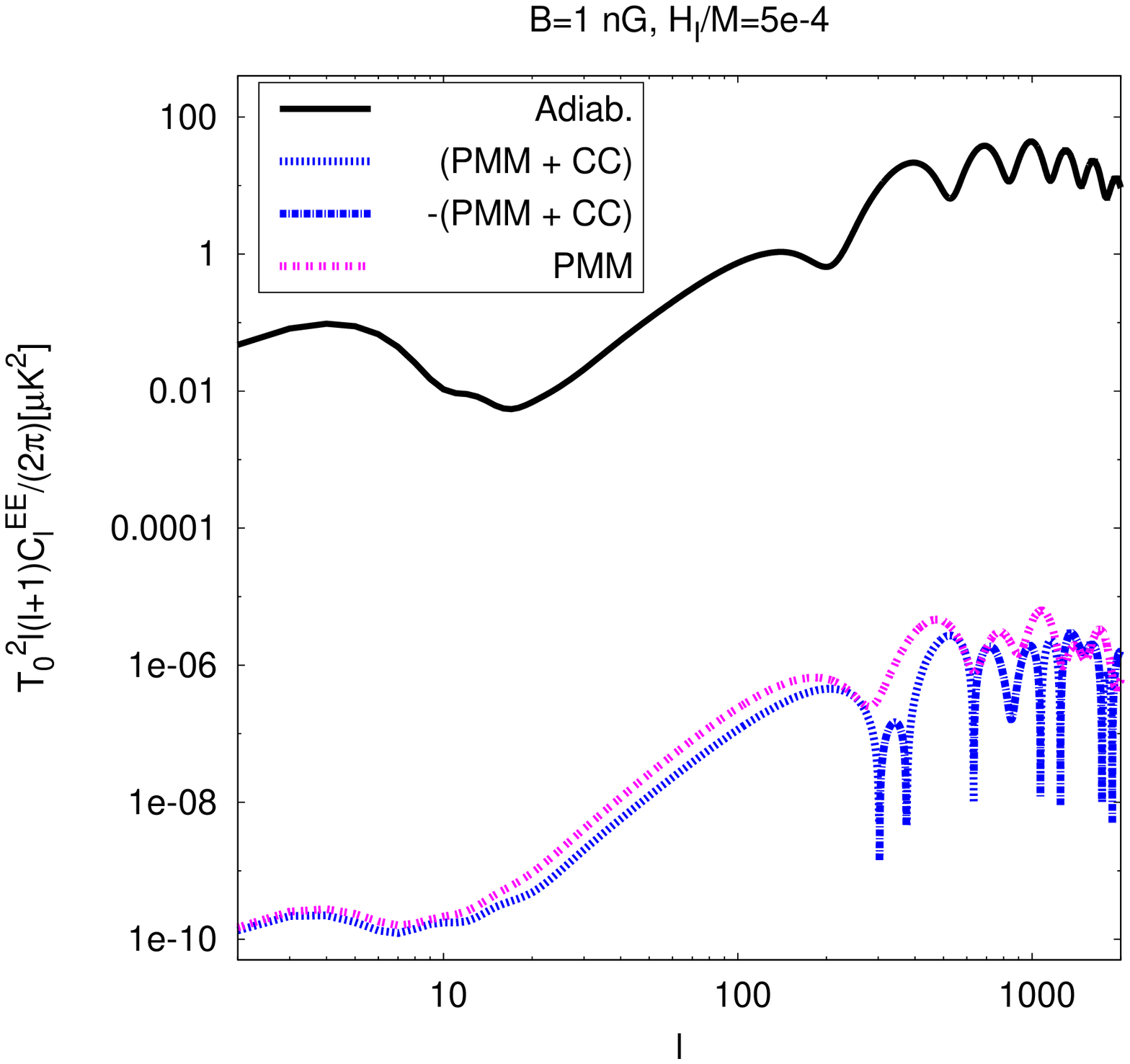}
}
\caption{Angular power spectra determining the autocorrelation of the polarization E-mode of the CMB
due to the cross correlation between the curvature mode $\zeta$ and the $\Delta_B$ mode   and 
due to the cross correlation between the curvature mode and the $\pi_B$ mode ({\it left}).
The contributions to the total angular power spectrum include the adiabatic mode ({\it Adiab.}), the pure magnetic mode ({\it PMM}) and the 
total cross-correlated magnetic-curvature mode ({\it CC}) ({\it right}). These have been calculated for WMAP9 $\Lambda$CDM
best fit parameters.
}
\label{fig4}
\end{figure}
Similarly, the contribution due to the curvature-magnetic modes cross correlation dominates at low multipoles  over the one due to the compensated 
magnetic mode for the polarization autocorrelation (cf. figure \ref{fig4}) as well as the temperature polarization cross correlation (cf. figure {\ref{fig5}). 
\begin{figure}[h!]
\centerline{\epsfxsize=3in\epsfbox{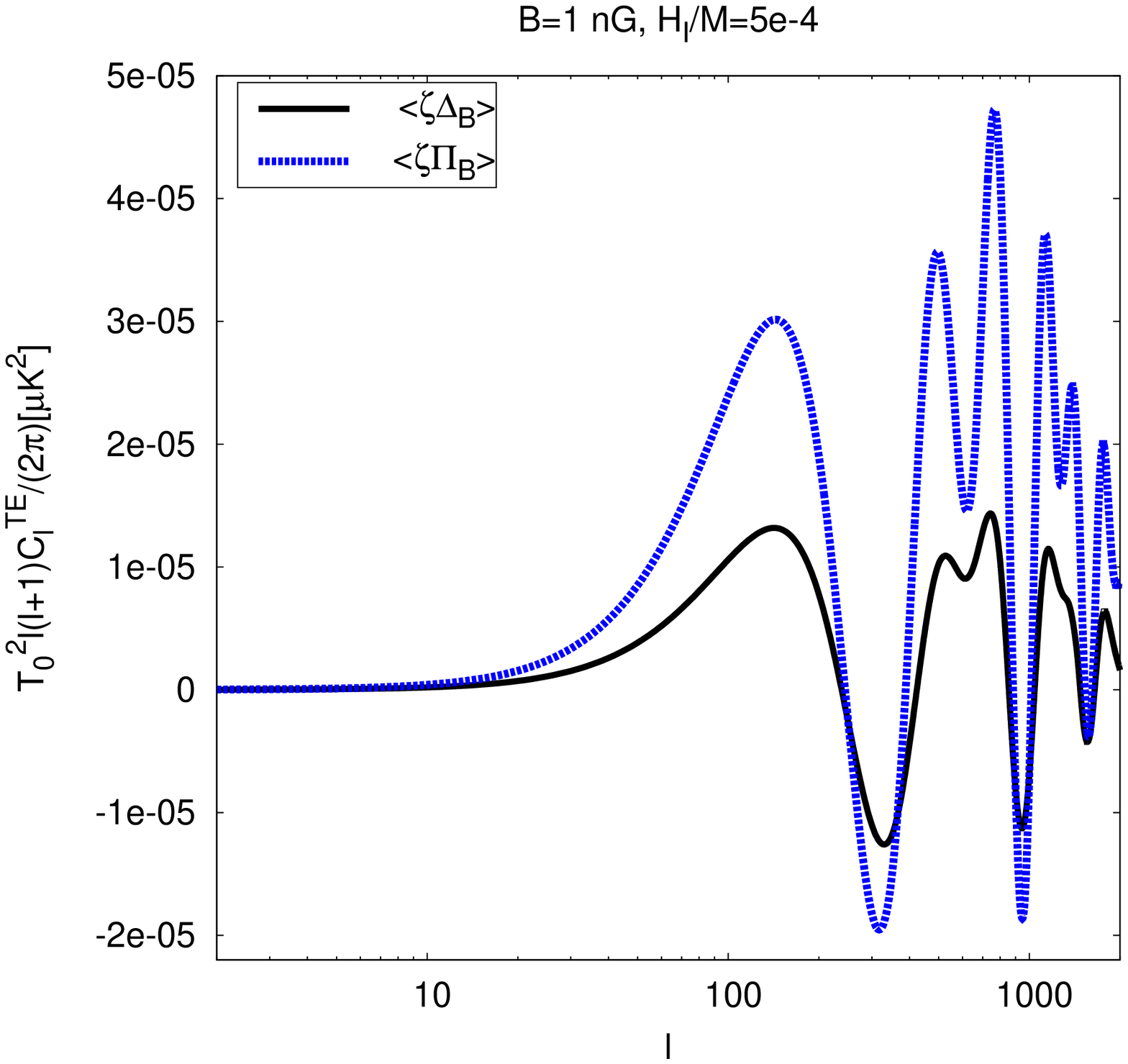}
\epsfxsize=3in\epsfbox{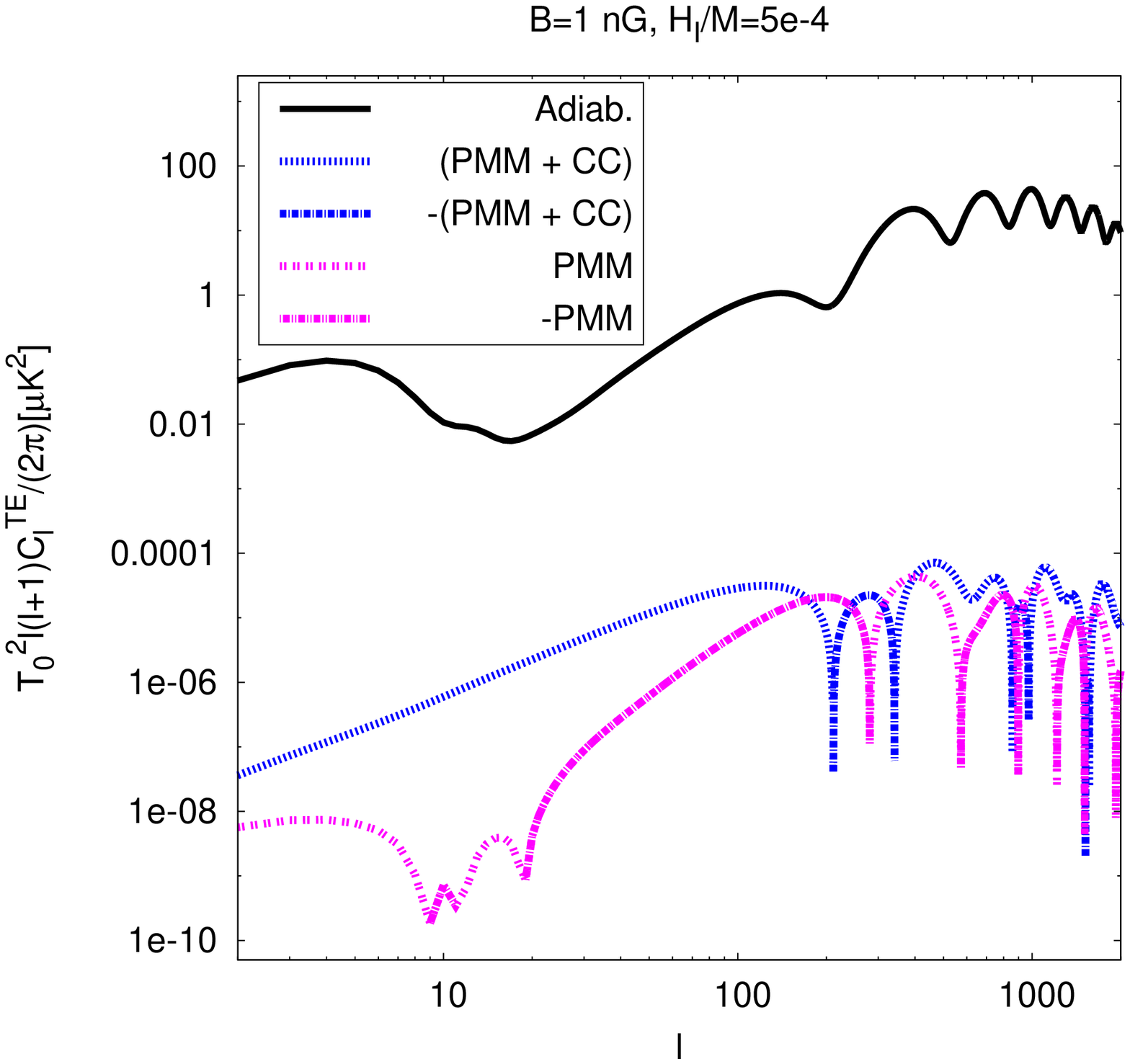}
}
\caption{Angular power spectra determining the cross correlation of temperature and polarization E-mode of the CMB
due to the cross correlation between the curvature mode $\zeta$ and the $\Delta_B$ mode  and 
due to the cross correlation between the curvature mode and the $\pi_B$ mode ({\it left}).
The contributions to the total angular power spectrum include the adiabatic mode ({\it Adiab.}), the pure magnetic mode ({\it PMM}) and the 
total cross-correlated magnetic-curvature mode ({\it CC}) ({\it right}). These have been calculated for WMAP9 $\Lambda$CDM
best fit parameters.
}
\label{fig5}
\end{figure}
In figure {\ref{fig6} the linear matter power spectrum is reported. On large scales the contribution due to the curvature-magnetic cross correlation
dominates over that of the pure magnetic mode. On small scales the former contribution becomes subleading resulting in the 
characteristic rise of the linear matter power spectrum in the presence of a magnetic field. This is due to the presence of the Lorentz term in the 
baryon velocity equation which on small scales dominates \cite{le,kk1}. Thereby leading to a local maximum on small scales in the total linear 
matter power spectrum taking into account all contributions, that is the adiabatic  and magnetic modes and their cross correlations. 
For wave numbers much larger than the maximal wave number $k_m$ determined by the damping of the magnetic field the linear matter 
power spectrum rapidly decays. In the case under consideration here $k_m=198$ Mpc$^{-1}$ (cf. equation (\ref{km})).
\begin{figure}[h!]
\centerline{\epsfxsize=3in\epsfbox{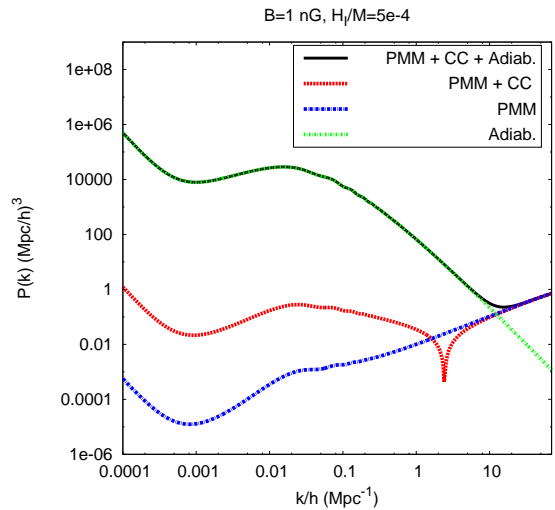}}
\caption{The total linear matter power spectrum  due to the pure magnetic modes ({\it PMM}),
the cross correlation bwetween the curvature mode and the magnetic modes ({\it CC}) and the pure adiabatic mode ($B=0$) ({\it Adiab.}) for the best-fit $\Lambda$CDM model of WMAP9.} 
\label{fig6}
\end{figure}

\section{Conclusions}
\setcounter{equation}{0}

Observations of the CMB of unprecedented precision allow to test  physics upto very early times. 
Observational evidence of large scale magnetic fields is  abundant. There is however the open question of their 
origin. Assuming magnetic fields have been created in the very early universe, long before decoupling, 
they influence the spectra of the anisotropies and polarization of the CMB as well the linear matter power spectrum.
There is clear observational evidence in the CMB pointing to the presence of an adiabatic mode which could be 
a primordial curvature mode generated during inflation (cf. e.g. \cite{wmap9}).
In general it has been assumed that the curvature mode and the magnetic field modes are uncorrelated.
However, even at lowest order, that is not taking into account nonlinear effects due to the evolution of the magnetic field \cite{evolv},
it seems natural to assume that non vanishing correlations exist depending on the particular model of magnetic field
generation during inflation, such as the coupling of electrodynamics to the inflaton \cite{mc,ssy} or, as the case considered here, to a spectator field
which is identified with the curvaton \cite{cmk}.
In \cite{cmk} the bispectrum determining the cross correlation between the magnetic field and a scalar field  
was calculated. Identifying the scalar field with the curvaton here the cross correlations between the curvature and the magnetic modes is
determined. Taking into account that the magnetic modes in the CMB are proportional to the magnetic field energy density and
 the magnetic anisotropic stress, respectively, both of which are quadratic in the magnetic field, the bispectrum or three-point function calculated in \cite{cmk} induces a spectrum or two-point function relevant for the calculation of the CMB and matter power spectrum.
The angular power spectra of the temperature anisotropies and polarization of the CMB have been calculated for a nearly scale invariant field
with field strength of 1 nG. It has been found that the contribution due to the cross correlation between the curvature mode and the magnetic modes
dominate at low multipoles over the contribution due to the pure magnetic mode. 
Similarly, the curvature magnetic mode cross correlated contribution dominates on large scales over the pure magnetic mode in the linear matter power spectrum. 
While on the contrary,  on small scales the pure magnetic mode dominates, not only over the cross correlated contribution but also the adiabatic contribution which is due to the presence of the Lorentz term in the baryon velocity equation.

\section{Acknowledgements}
I would like to thank the CERN Theory Division as well as the Max-Planck-Institute for Astrophysics for hospitality where part of this work was done.
Financial support by Spanish Science Ministry grants FPA2009-10612, FIS2012-30926 and CSD2007-00042 is gratefully acknowledged.
Furthermore I  acknowledge the use of the Legacy Archive for Microwave Background Data Analysis (LAMBDA). Support for LAMBDA is provided by the NASA Office of Space Science.

\end{document}